# Extracting intrinsic superconducting properties in intercalated layered superconductors using an extended 2D Tinkham model

Yue Liu,[1,2,†] Yuhang Zhang,[1,2,†] Zouyouwei Lu,[1,2,†] Dong Li,[1,3,*] Yuki M. Itahashi,[3] Zhanyi Zhao,[1,2] Jiali Liu,[1,2] Jihu Lu,[1,2] Feng Wu,[1,4] Kui Jin,[1,2,5] Hua Zhang,[1] Ziyi Liu,[1] Xiaoli Dong,[1,2,5,**] Zhongxian Zhao,[1,2,5]

[1]Beijing National Laboratory for Condensed Matter Physics, Institute of Physics, Chinese Academy of Sciences, Beijing 100190, China.
[2]School of Physical Sciences, University of Chinese Academy of Sciences, Beijing 100049, China.
[3]RIKEN Center for Emergent Matter Science (CEMS), Saitama 351-0198, Japan.
[4]Key Laboratory of Advanced Optoelectronic Quantum Architecture and Measurement, Ministry of Education, School of Physics, Beijing Institute of Technology, Beijing 100081, China.
[5]Songshan Lake Materials Laboratory, Dongguan, Guangdong 523808, China.

**ABSTRACT**. Bulk two-dimensional (2D) superconductivity has gained considerable attention due to its intricate interplay between symmetry breaking, nontrivial topology, 2D phase fluctuations, and unconventional superconductivity. However, certain intercalated layered superconductors, despite their short $c$-axis superconducting coherence length, have been misclassified as anisotropic three-dimensional (3D) superconductors. Here, we investigate (Li,Fe)OHFeSe superconductors with varying degrees of interlayer misalignment, revealing sample-dependent superconducting dimensionality while consistently observing Berezinskii–Kosterlitz–Thouless (BKT) transitions. To resolve this discrepancy, we develop an extended 2D Tinkham model that quantitatively captures the blurring effects induced by interlayer misalignment. We further demonstrate the validity of this model in both (Li,Fe)OHFeSe and cetyltrimethyl ammonium (CTA$^+$)-intercalated (CTA)$_{0.5}$SnSe$_2$ superconductors, highlighting its broad applicability. This work provides valuable insights into bulk 2D superconductivity and establishes an extended 2D Tinkham model for quantitatively extracting intrinsic superconducting properties in intercalated layered superconductors, particularly those exhibiting significant interlayer misalignments.

† These authors contributed equally.
* Contact author: dong.li.hs@riken.jp
** Contact author: dong@iphy.ac.cn

## I. INTRODUCTION.

Advances in two-dimensional (2D) superconductivity have attracted significant attention due to the pursuit of long-standing predictions of exotic Cooper pairing [1-9]. All these unconventional superconductivities stem from the intricate interplay between Cooper pairs and the surrounding structural environments. For instance, in conventional three-dimensional (3D) superconductors, Cooper pairs condense and become coherent simultaneously, with most properties well described by the Bardeen–Cooper–Schrieffer (BCS) theory [10]. In contrast, in layered superconductors, strong fluctuation effects inherent to their 2D nature allow the formation of preformed Cooper pairs above the critical temperature $T_c$, preceding the establishment of long-range superconducting coherence [11]. Moreover, the fragile finite-momentum pairing [5-7] and superconducting stripes [8, 12] may persist due to spatial modulations that optimize energy costs, rendering them more favorable than conventional Cooper pairs. These puzzling phenomena have boosted extensive investigation into 2D superconductivity.

2D superconductivity can be realized in two types of platforms, thin films and bulk samples, via different mechanisms. Owing to the rapid advancements in thin film growth and 2D device fabrication, thin samples host 2D superconductivity when the thickness is less than twice the out-of-plane superconducting coherence length $\xi_\perp$ [1]. Representative materials include the SrTiO$_3$/LaAlO$_3$ interface [13], monolayer FeSe [14], gated MoS$_2$ [15, 16], and exfoliated few-layer NbSe$_2$ [17], etc, where the excitations of in-plane vortices are effectively forbidden. An alternative method to stabilizing (quasi-) 2D superconductivity is to expand the $c$-axis lattice parameter $c$ by intercalating non-superconducting functional layers in bulk samples. When $c > \xi_\perp$, weak Josephson interlayer coupling emerges between adjacent superconducting layers, replacing long-range superconducting coherence in the bulk [18-21]. The latter does not require any physical constraints on the sample thickness and exhibit the intrinsic Josephson effects, as observed in cuprates like $Bi_2Sr_2CaCu_2O_{8+\delta}$ [19] and iron-based superconductors such as $(V_2Sr_4O_6)Fe_2As_2$ [21]. Meanwhile, the intercalated layers could introduce additional exotic properties, such as nontrivial topological electronic bands [22], chirality [23], and inversion-symmetry breaking [5], further enhancing the significance of investigating 2D superconductivity in bulk samples.

However, layered superconductors are often mistakenly classified as exhibiting anisotropic three-dimensional (3D) superconductivity when the condition of $c > \xi_\perp$ is satisfied. For examples, while most layered superconductors, such as the cuprate $Bi_2Sr_2CaCu_2O_{8+\delta}$ [24] and misfit transition metal dichalcogenides (TMDs) compound $Ba_6Nb_{11}S_{28}$ [5], follow the expected 2D superconductivity, others, including the organic molecule intercalated $(CTA)_{0.5}SnSe_2$ [25], iron pnictide $KCa_2Fe_4As_4F_2$ [26], and iron chalcogenide (Li,Fe)OHFeSe [27] deviate from the 2D anticipation despite owing $\xi_\perp$ values shorter than the corresponding $c$ lengths. Especially, the intercalated FeSe superconductor (Li,Fe)OHFeSe has been reported to exhibit conflicting dimensional characteristics, with some studies suggesting anisotropic 3D behavior [28, 29] while others indicate quasi-2D superconductivity [30-32]. These contradictory results motivated us to scrutinize the dimensional nature of (Li,Fe)OHFeSe, serving as a prototype of intercalated layered superconductors.

In this letter, we investigated (Li,Fe)OHFeSe single crystals and epitaxial films, revealing that interlayer misalignment significantly affects the measured superconducting properties in the intercalated layered superconductors. The (Li,Fe)OHFeSe epitaxial films exhibit improved $c$-axis orientation, as evidenced by the full width at half maximum (FWHM) values of (006) peak reduced from 1.49° in single crystals to 0.12°. This distinction in interlayer alignment results in sample-dependent angular dependence of upper critical field $H_{c2}(\theta)$ behaviors between (Li,Fe)OHFeSe epitaxial films and single crystals. The single crystals appear to exhibit anisotropic 3D superconductivity, while the films show 2D superconductivity, despite both displaying BKT transitions. To quantify this discrepancy, we incorporated the measured FWHM values into an extended 2D Tinkham model via a convolution function, quantitatively reproducing experimental results with more reasonable fitting parameters. Furthermore, based on the extended model we developed, we confirmed the bulk 2D superconductivity nature in the $(CTA)_{0.5}SnSe_2$ superconductor, which was poorly fitted by the standard 2D Tinkham model without accounting for significant interlayer misalignment issues. This work provides a simple yet effective approach for extracting intrinsic superconducting properties in intercalated layered superconductors, particularly those synthesized via soft chemical methods.

## II. METHODS

The (Li,Fe)OHFeSe single crystals were synthesized by the hydrothermal ion-exchange method we pioneered [33, 34], and the (Li,Fe)OHFeSe epitaxial films were grown on $LaAlO_3$ substrates via the so-called matrix-assisted hydrothermal epitaxy (MHE) we developed [34-36]. X-ray diffraction (XRD) data of all the studied samples were collected at room temperature on a 9 kW Rigaku SmartLab x-ray diffractometer equipped with two Ge(220) monochromators. The measured samples have thicknesses of 600–700 nm for the films (determined on a Hitachi SU5000 scanning electron microscope) and of about 80 microns for the single crystals, with the lateral size of millimeters for both the samples. The electrical transport properties were measured on a Quantum Design PPMS-9 system with the standard four-probe method. During the angle-dependent measurements, the samples were rotated to change the tilting angle $\theta$ between the magnetic field $B$ and crystallographic $c$-axis using a Quantum Design high resolution rotator. The DC current-voltage ($I$-$V$) measurements were collected by two Keithley meters 6221 and 2182, similar to previous reports [30, 37]. The upper limits of applied current are 5 mA and 20 mA in the measurements of $I$-$V$ curves for (Li,Fe)OHFeSe films and crystals, respectively.

## III. RESULTS AND DISCUSSION

Fig. 1(a) illustrates the X-ray rocking curve measurements, which evaluate crystalline orientation quality based on the FWHM values. The (Li,Fe)OHFeSe single crystals are synthesized by intercalating (Li,Fe)OH molecules into FeSe layers [33, 34], resulting in a large $c$-axis lattice parameter of about 9.3Å, but with imperfect crystalline orientation along the interlayer direction. Although the XRD patterns exhibit sharp diffraction peaks, confirming the single-crystalline nature of (Li,Fe)OHFeSe, the FWHM value of (006) diffraction peak is relatively large at 1.49° for the bulk crystals (Fig. 1(b)). To further enhance crystalline quality, particularly interlayer alignment, we developed a new synthesis method by introducing the LaAlO$_3$ substrates to grow (Li,Fe)OHFeSe thick films [35, 38]. The in-plane lattice mismatch is negligible between LaAlO$_3$ ($a \sim$ 3.787Å) and (Li,Fe)OHFeSe ($a \sim$ 3.78Å), enabling the high-quality epitaxial growth [22, 35]. Consequently, the corresponding FWHM value is reduced by an order of magnitude to 0.12° in the epitaxial films (Fig. 1(c)). Such a noticeable difference in crystalline quality may account for the distinct observations reported in previous studies for these two types of samples. Specifically, pair-breaking peaks and collective modes in electronic Raman spectroscopy [39], zero-energy vortex modes in scanning tunneling spectroscopy/microscopy (STS/M) [22, 40], and the spatial distributions of vortex bound states [41] have been observed exclusively in (Li,Fe)OHFeSe epitaxial films, whereas they are absent in (Li,Fe)OHFeSe single crystals when examined using the same techniques [42, 43]. These distinct observations in micro-spectroscopy highlight that improved crystalline quality enables the accurate probing of intrinsic properties in intercalated layered superconductors. Therefore, we further compare their electrical transport properties to gain insight into their macroscopic bulk characteristics, especially regarding the superconducting properties.

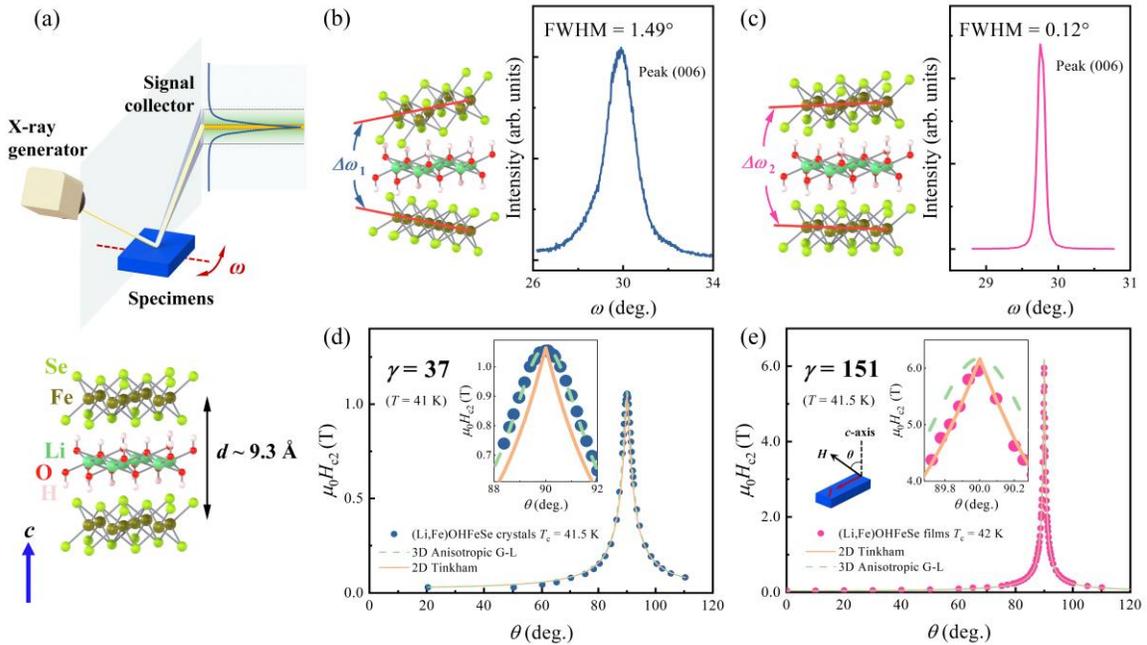

FIG 1. Interlayer misalignment induced sample-dependent results. (a) The schematic illustrations for the measurement geometry of XRD rocking curve and the crystal structure of (Li,Fe)OHFeSe. (b) and (c) X-ray rocking curves for the (006) peaks of (Li,Fe)OHFeSe single crystals and epitaxial films, respectively. A larger full width at half maximum (FWHM) value indicates worse interlayer misalignment, as shown in the insets, with the $\Delta\omega_1$ and $\Delta\omega_2$ values exaggerated for clarity. (d) and (e) Angular dependence of the upper critical field $\mu_0H_{c2}$ extracted at 10% $R_n$ for (Li,Fe)OHFeSe single crystals (at $T$ = 41.0 K) and epitaxial films (at $T$ = 41.5 K), respectively. The light orange and olive curves represent the fittings of the 2D Tinkham model and the 3D anisotropic G-L model, respectively. And the inset of Fig. 1(e) illustrates the angular rotation ($\theta$) in electrical transport.

A common method for probing the nature of superconducting dimensionality is to analyze the angular dependence of the upper critical field $H_{c2}(\theta)$, by fitting it to either the 3D anisotropic Ginzburg-Landau (G-L) model, $(H_{c2}(\theta)\cos\theta/H_{c2}^c)^2 + (H_{c2}(\theta)\sin\theta/H_{c2}^{ab})^2 = 1$, or the 2D Tinkham model, $|H_{c2}(\theta)\cos\theta/H_{c2}^c| + (H_{c2}(\theta)\sin\theta/H_{c2}^{ab})^2 = 1$ [44]. We measured the magnetic-field-dependent resistance at different tilting angles from $\theta$ = 0° ($H$ // $c$) to $\theta$ = 90° ($H$ // $ab$) for both the (Li,Fe)OHFeSe single crystals and epitaxial films. The corresponding $H_{c2}$ values were extracted from the 10% normal-state resistance $R_n$, defined as $R_{50\,K,\,0\,T}$. Figs. 1(d) and 1(e) show the $H_{c2}(\theta)$ data of (Li,Fe)OHFeSe single crystals and epitaxial films, respectively. We attempted to fit their $H_{c2}(\theta)$ line shapes by both the 2D Tinkham model and the 3D anisotropic G-L model. Remarkably, the $H_{c2}(\theta)$ dots closely follow the 3D anisotropic G-L model for the (Li,Fe)OHFeSe single crystals, while they are well fitted by the 2D Tinkham model for the (Li,Fe)OHFeSe epitaxial films, as indicated by the distinct rounded shape and cusp-like peak near $\theta = 90°$. And these fittings yield significantly different superconducting anisotropy values, $\gamma = H_{c2}^{ab}/H_{c2}^{c}$, for the (Li,Fe)OHFeSe single crystals ($\gamma = 37$) and epitaxial films ($\gamma = 151$). It is noted that the intermediate angle range is poorly fitted in the former, while the entire angle range is well fitted in the latter, indicating that the 2D feature observed in the epitaxial films provides a more reliable result for (Li,Fe)OHFeSe superconductors. We emphasize that the observed 2D feature stems from intrinsic bulk 2D superconductivity, rather than being induced by the thin-film geometry, as the thickness of the (Li,Fe)OHFeSe films (600-700 nm) is significantly larger than $2\xi_\perp = 0.48$ nm [30]. Therefore, we speculate that similar bulk 2D superconductivity also occurs in the (Li,Fe)OHFeSe single crystals.

The excitations of bound vortex-antivortex pairs are anticipated in bulk 2D superconductivity under zero magnetic field, which account for the widely observed Berezinskii–Kosterlitz–Thouless (BKT) transitions in various 2D superconductors [1, 5, 45]. Experimentally, two key characteristics of BKT transitions are manifested: the scaled resistive transition and the jumped exponent $\alpha$ in current-voltage (I-V) curves, where $V \propto I^\alpha$. Both features were successfully observed in the (Li,Fe)OHFeSe single crystals and epitaxial films, as shown in Fig. 2, confirming the presence of intrinsic bulk 2D superconductivity in (Li,Fe)OHFeSe superconductors, as the condition $c > \xi_\perp$ is satisfied. The linear Halperin–Nelson scaling law, $(d\ln R/dT)^{-2/3}$ vs. T, yield the BKT transition temperature $T_{BKT} = 41.2$ K and 42.1 K for the single crystals and epitaxial films, respectively (see the upper panel in Fig.2(c)). These values are consistent with the fitted $T_{BKT} = 41.3$ K and 42.0 K at $\alpha(T_{BKT}) = 3$ from I-V characteristics (see the lower panel in Fig.2(c)). It is noted that the $T_c = 41.5$ K (Fig.1) and 41.0 K (Fig.2) of the single crystals we studied are slightly lower than those of the epitaxial films ($T_c = 42.0$ K) due to the stoichiometry issues, leading to a lower $T_{BKT}$ in single crystals. Specifically, the (Li,Fe)OHFe$_{1-x}$Se single crystal contains a higher iron vacancy concentration ($x \sim 2\%$) in the Fe$_{1-x}$Se layers [33] compared to $x = 0.54\%$ in the (Li,Fe)OHFe$_{1-x}$Se thin films [46]. However, these discrepancies cannot explain the distinct $H_{c2}(\theta)$ evolutions observed in Fig. 1, as iron vacancies primarily suppress superconductivity by promoting the in-plane competing charge orders as $x$ values from 0.54% to 6.35% [46]. Therefore, we will omit this negligible stoichiometric deviation and the resulting $T_c$ difference across different samples in the following discussions, as these factors do not affect our main conclusions and fall beyond the scope of this study.

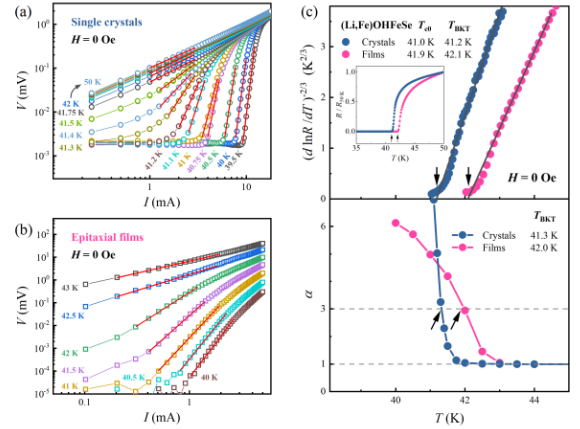

FIG 2. BKT transitions in both (Li,Fe)OHFeSe single crystals and epitaxial films. (a) and (b) Current-voltage (I-V) curves of (Li,Fe)OHFeSe single crystals and epitaxial films measured at different temperatures, respectively. The red lines are the power-law ($V \propto I^\alpha$) fits to the curves. (c) The upper panel shows the resistive transitions of (Li,Fe)OHFeSe crystals (navy blue) and films (pink) plotted as $(d\ln R/dT)^{-2/3}$ vs. T. And the inset is temperature dependence of the normalized resistivity $R/R_{50\ K}$ for (Li,Fe)OHFeSe crystals and films with zero resistance temperature $T_{c0} \sim 41.0$ K and 41.9 K, respectively. The lower panel shows temperature dependence of the power-law fitted exponent $\alpha$ for (Li,Fe)OHFeSe crystals and films, respectively.

Given that BKT transitions are confirmed in both samples, the intrinsic 2D-type $H_{c2}(\theta)$ evolution is accurately captured in (Li,Fe)OHFeSe epitaxial films, whereas in (Li,Fe)OHFeSe single crystals, it is obscured by additional factors. The dominating difference between these two samples is the FWHM values of (006) diffraction peaks, primarily related to the interlayer alignment along the c-axis (Fig. 1(b) and (c)). More technically, the FWHM values are the integrated results including the strain effect [47], instrumental broadening [48], crystalline defects or impurities [49], etc. Since the contributions from these non-crystalline factors like instrumental broadening should be constant and smaller than 0.12°, the smallest FWHM value we observed, the significant broadening of (006) diffraction peaks in (Li,Fe)OHFeSe single crystals are attributed to crystalline factors. In the geometric configuration of (00l) rocking curve measurements, interlayer misalignments are the dominant source of broadening (Fig. 1(a-c)). We thereby consider the effect of c-axis misalignment on the magnetoresistance $R(B)$ under different tilting angle ($\theta$) we measured, as both the X-ray incident angle ($\omega$) and field titling angle ($\theta$) rotate within the same coplanar plane. As a result, the definition of $\theta$ becomes blurred by the sample-dependent FWHM (or $\Delta\omega$) values, which complicates the interpretation of $R(B, \theta)$ and the precise extraction $H_{c2}(\theta)$ data. On the other hand, the imperfect $c$-axis alignment may distort the cusp-like profile of $H_{c2}(\theta)$ near $\theta = 90°$, leading to an underestimation of the anisotropy $\gamma$ value as we observed in (Li,Fe)OHFeSe single crystals.

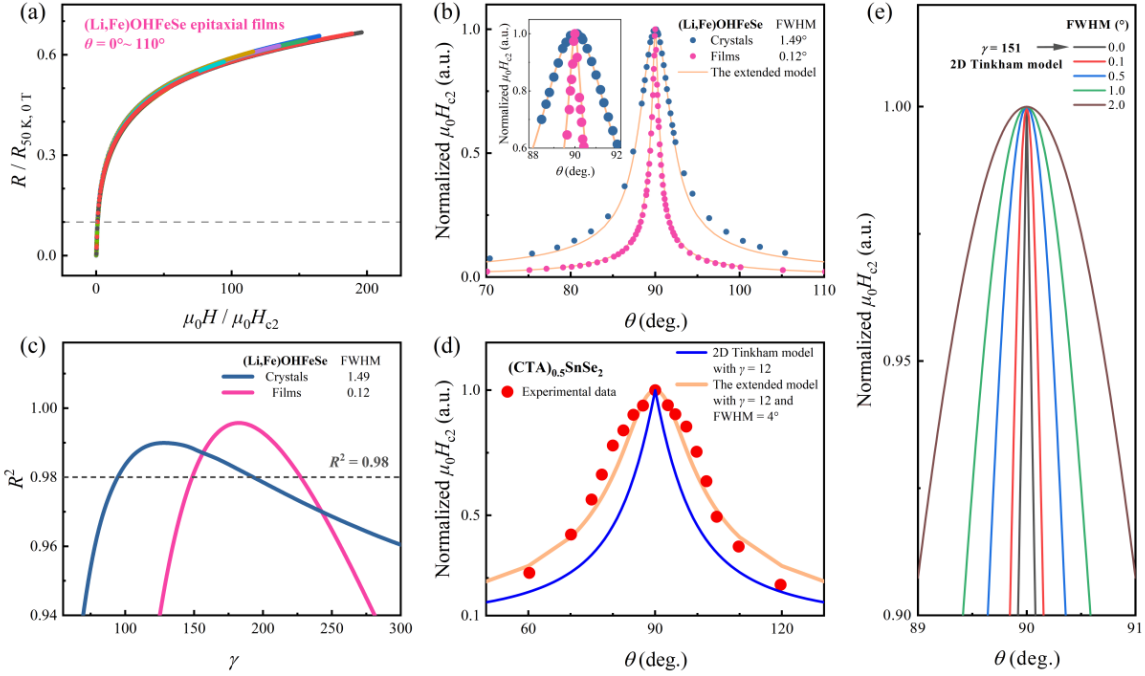

FIG 3. Extracting the intrinsic superconducting properties in intercalated layered superconductors via an extended 2D Tinkham model. (a) Scaling behavior of the normalized resistance $R/R_{50\,K}$ ($\mu_0H$, $\theta$) versus $\mu_0H / \mu_0H_{c2}$ at different tilting angles for (Li,Fe)OHFeSe epitaxial film. (b) The fitted $\mu_0H_{c2}(\theta)$ curves for (Li,Fe)OHFeSe single crystals (navy blue) and epitaxial films (pink) using the extended model. The anisotropy $\gamma$ is fixed at 151, and the measured FWHM of 1.49° (single crystals) and 0.12° (epitaxial films) are incorporated into the fitting process. (c) The fitting goodness parameters for the $\mu_0H_{c2}(\theta)$ curves with varying $\gamma$ values in both (Li,Fe)OHFeSe single crystals and epitaxial films. $R^2$ is derived from the difference between experimental data and the extended 2D Tinkham model fitting. (d) Comparison of the fits between the fitted $\mu_0H_{c2}(\theta)$ curves using both the extended (orange) and standard (blue) 2D Tinkham models with $\gamma$ fixed at 12. The FWHM = 4° and the experimental data for $(CTA)_{0.5}SnSe_2$ are both extracted from the reference [25]. (e) The $\mu_0H_{c2}(\theta)$ curves described by the extended 2D Tinkham model with $\gamma = 151$ and increased FWHM values. The standard 2D Tinkham model corresponds to the curve with an FWHM of 0°.

To extract the intrinsic superconducting properties from the sample-dependent $H_{c2}(\theta)$ data, we propose an extended method based on the 2D Tinkham model. Compared with the ideal crystals described by the theorical model, the simulation process considers the interlayer alignment of the practical samples. Firstly, we find that the normalized magnetoresistance $R/R_{50\,K}$ ($\mu_0H$, $\theta$) at different tilting angles for both (Li,Fe)OHFeSe epitaxial films and single crystals can be approximatively scaled into a single curve, as represented by the data of films in Fig. 3(a). The scaling process can be expressed as $R$ ($\mu_0H$, $\theta$) = $R$ ($h$), where $h = \mu_0H / \mu_0H_{c2}(\theta)$, which was theoretically proposed in layered high-$T_c$ superconductors [50]. This scaling demonstrates that the superconducting transition in (Li,Fe)OHFeSe exhibits the same nature, regardless of the detailed sample quality. Next, we modeled the macroscopic resistance as the result of the series connection of each layer, with interlayer mosaics quantitatively determined from the obtained FWHM ($\Delta\omega$) values from rocking curve measurements. Therefore, by incorporating the sample-dependent $\Delta\omega$ values into the ideal 2D Tinkham model ($\Delta\omega = 0$), the measured resistance $R^{measure}(H,\theta)$ can be corrected through a simple convolution of $R$ and $\Delta\omega$, which can be expressed as

$$R^{measure}(H,\theta) = \int g(\omega) \cdot R(H, \theta - \omega)\, d\omega$$
$$= \int g(\omega) \cdot R\left(\frac{H}{H_{c2}(\theta-\omega)}\right) d\omega \quad (1)$$

where the $g(\omega)$ is the Gaussian distribution with the sample-dependent FWHM ($\Delta\omega$) values, $R(h)$ is from the scaled resistance in Fig. 3(a), and $H_{c2}(\theta)$ corresponds to the 2D Tinkham model with assumed $\gamma = 151$. We use the Gauss function to statistically reflect the interlayer misalignments, like the methods employed in previous literatures [51, 52]. We then extract $H_{c2}^{measure}(\theta)$ using the same 10% $R^{measure}(H,\theta)$ criterion.

Based on the extended 2D-Tinkham model proposed in Eq. (1), we successfully fitted the measured $H_{c2}(\theta)$ data of (Li,Fe)OHFeSe single crystals and epitaxial films, using the $\gamma$ value fixed at 151 and sample-dependent FWHM values of 0.12° and 1.49°, respectively (Fig. 3(b)). To determine the intrinsic $\gamma$ values and minimize the influence of other factors, such as the motions of out-of-plane pancake vortices [30], we restricted the fitting of the $H_{c2}(\theta)$ data to the range near 90° (from 80° to 100°). The fitting goodness parameters, $R^2$, as a function of $\gamma$ values are shown in Fig. 3(c), yielding $\gamma$ values of 95 ~ 193 and 149 ~ 227 for the (Li,Fe)OHFeSe single crystals and epitaxial films, respectively, when $R^2 > 0.98$. Although the fitted $\gamma$ values show slight differences across different samples in our toy model, the best fit $\gamma = 128$ for single crystals via the extended 2D Tinkham model is more reasonable than the $\gamma = 37$ obtained from the 3D anisotropic G-L model, thereby confirming the advantage of the extended 2D Tinkham model in extracting intrinsic superconducting anisotropy parameter.

To assess the universal applicability of the extended 2D Tinkham model beyond (Li,Fe)OHFeSe, we applied it to other molecule-intercalated layered superconductors like the $(CTA)_{0.5}SnSe_2$ [25]. While BKT transitions were observed in $(CTA)_{0.5}SnSe_2$ [25], the $H_{c2}(\theta)$ data exhibited significant deviations from the standard 2D Tinkham model. One notable feature is that $(CTA)_{0.5}SnSe_2$ exhibits a large FWHM value of 4° for the (001) diffraction peak, which is comparable to that of (Li,Fe)OHFeSe single crystals. We used the experimentally measured parameters, FWHM = 4° and $\gamma = 12$, to fit the $H_{c2}(\theta)$ line shape using both the extended (light orange) and standard (blue) 2D Tinkham model, as shown in Fig. 3(d). The extended model provides a much better fit compared to the standard model, without introducing any additional degrees of freedom. This demonstration not only verifies the bulk 2D superconductivity nature of $(CTA)_{0.5}SnSe_2$, an issue not fully elucidated in the previous report [25], but also highlights the robustness and broad applicability of the extended 2D Tinkham model in analyzing layered superconductors. To clearly illustrate the broadening effect of sample-dependent FWHM values on the $H_{c2}(\theta)$ line shapes, we simulated the $H_{c2}(\theta)$ curves with varying FWHM ranging from 0° to 2° using the extended 2D Tinkham method (Fig. 3(e)). As the FWHM values increase, the peak shapes gradually transition from the cusp-like (standard model) form to a more rounded one (extended model). Therefore, it is essential to take the practical interlayer misalignments into account when analyzing the $H_{c2}(\theta)$ data of layer superconductors, particularly for the intercalated compounds with broad interlayer mosaics.

We envision broad applications for this extended 2D Tinkham model on various interlayered superconductors with imperfect interlayer alignments, a common feature in the intercalated compounds [25, 53] and frequently encountered during the early stages of exploring new superconductors [54, 55]. From the perspective of sample synthesis, conventional solid-state reactions involve an annealing process at high temperatures, allowing atoms to form the perfect long-range crystal lattices, as observed in materials like $PbTaSe_2$ [56, 57] or $(PbSe)_{1.14}NbSe_2$ [58]. On the contrast, to incorporate various functional molecules that are generally unstable at high temperatures, soft chemical methods are often employed at relatively low temperatures, at the expense of abandoning high-temperature annealing process. Therefore, soft chemical methods, such as the hydrothermal methods [33, 34, 53] or the widely used electrochemical intercalation [25, 59-61], inevitably introduce crystalline imperfections related to interlayer misalignments. The adverse effects of imperfect crystalline quality have been suggested in previous works from three decades ago to nowadays [25, 52, 62], but likely due to the lacking of a comparative system with tunable FWHM values, solid experimental evidence or reliable theoretical models have not been reported until this work. Our results provide valuable insights and a universal method for investigating bulk 2D superconductivity in intercalated layered materials, particularly concerning interlayer misalignment issues.

## IV. CONCLUSIONS

In summary, we comparatively investigated the (Li,Fe)OHFeSe superconductors with varying FWHM values and proposed an extended 2D Tinkham model to extract the intrinsic superconducting properties of intercalated layered superconductors. The FWHM of the (006) peaks in (Li,Fe)OHFeSe was reduced by an order of magnitude, from 1.49° in single crystals to 0.12° in thick films due to epitaxial growth. Remarkably, while BKT transitions were consistently observed in both samples, the $H_{c2}(\theta)$ data exhibited sample-dependent behavior. This discrepancy motivates us to incorporate interlayer misalignment into the standard 2D Tinkham model via a convolution function. The extended model successfully scaled the resistive curves and quantitatively fitted the experimental $H_{c2}(\theta)$ data for both single crystals and epitaxial films, yielding the more reasonable $\gamma$ fitting values. Moreover, the extended model also provided a better fit than the standard 2D Tinkham in the case of $(CTA)_{0.5}SnSe_2$, demonstrating its robustness and broad applicability in analyzing intercalated layered superconductors.

## ACKNOWLEDGMENTS

We acknowledge Fang Zhou and Yoshihiro Iwasa for their insightful comments and to Mari Ishida for her assistance in depicting schematic figures. This work was supported by the National Key Research and Development Program of China (Grant Nos. 2022YFA1403903, and 2023YFA1406101), the Strategic Priority Research Program of Chinese Academy of Sciences (XDB33010200), the National Natural Science Foundation of China (Grants No. 12304075) and CAS Project for Young Scientists in Basic Research (2022YSBR-048).

D. L. and X.L. D. conceived this project. Y. L., Z.Y.W. L., and D. L. synthesized and characterized the samples under supervision from X.L. D.. Y.H. Z. developed the extended 2D Tinkham Model and conducted the simulations. Y. L., D. L., Y.H. Z., Z.Y. Z., J.L. L., F. W. and Z.Y. L. performed the transport measurements. Y. L., Y.H. Z., D. L., Y.M. I., and X.L. D. analyzed the data. D. L., Y. L. and X.L. D. wrote the manuscript with contributions from all authors. X.L. D. and Z.X. Z. supervised the project.